# On a different expression of the classical limit of quantum mechanics.

Alejandro A. Hnilo

*CEILAP, Centro de Investigaciones en Láseres y Aplicaciones, UNIDEF (MINDEF-CONICET);*
*CITEDEF, J.B. de La Salle 4397, (1603) Villa Martelli, Argentina.*
email: ahnilo@citedef.gob.ar
October 2nd, 2018.

The principle of correspondence (or classical limit) is essential in quantum mechanics. Yet, how and why quantum phenomena vanish at the macroscopic scale are issues still open to debate. Here, quantum mechanical predictions for Greenberger-Horne-Zeilinger states of qubits are shown to be easier to reproduce with a classical model as the number of particles increases, even in the absence of loopholes or conspiratorial mechanisms of any kind. It is conjectured that this result may lead to the simplest way to express the principle of correspondence.

The *principle of correspondence* is a fundamental idea in Quantum Mechanics (QM). It states that results predicted by QM must converge to the ones predicted by Classical Mechanics as the macroscopic scale is reached. This condition is usually expressed in terms of a vanishing Planck's constant ($h \rightarrow 0$) as scaled with the system's action. In this limit, all operators commute and Hamilton's principle takes its classical expression. When dealing with movement of particles, classical trajectories are retrieved except for chaotic systems. In general, quantum effects as superposition, interference and entanglement are expected to vanish at the macroscopic scale. Yet, this vanishing is not easy to explain. The customary explanation involves some interaction with a Markovian environment where quantum coherence is shared among a huge number of degrees of freedom and, eventually, lost. Alternative explanations have been proposed. I just mention here non-Hermitian evolution in rigged Hilbert spaces [1], the assumption of new physics inducing a spontaneous collapse of the wavefunction [2] and the hypothesis that no decay actually occurs and that parallel realities exist (many-worlds interpretation) [3]. A review of alternative explanations can be found in [4].

In this short contribution, I point out the simple (but perhaps surprising) result that the classical limit is reached by merely increasing the number of elementary particles in the system being considered. In order to get this result, I make two assumptions:

*i)* That the principle of correspondence can be also stated in this way: ask whether the results of all observations on the system can be explained by a classical model, or not. If the answer is "yes", the classical limit is reached. This criterion seems to be more general than (and to include) the usual one. It is independent of the system's action and the existence of trajectories.

*ii)* That Greenberger-Horne-Zeilinger (GHZ) is the maximal form of entanglement among an arbitrary number of particles. In other words: that given a system of $q$ elementary particles, its state farthest from classical is a GHZ state. As far as I know, no unanimously accepted measure of entanglement for systems of many particles exists, but this assumption is often believed to be true.

I think the result to be presented here may be surprising because of the widespread belief that GHZ states "single shot" disprove Local Realism and that they are, in consequence, the utmost example of non-classicality. Let see the result now.

Following the usual approach [5], consider a GHZ state of $q$ qubits:

$$|\phi_{(q)}\rangle = 1/\sqrt{2}\{|x_1,...,x_q\rangle + i\,|y_1,...,y_q\rangle\} \qquad (1)$$

which is observed at $q$ stations spatially separated of each other. In each station a Pauli spin operator $\sigma_l$ or $\sigma_r$ is applied: $\sigma_l|x\rangle = |y\rangle$, $\sigma_l|y\rangle = |x\rangle$, $\sigma_r|x\rangle = i|y\rangle$, $\sigma_r|y\rangle = -i|x\rangle$. The operator is chosen randomly in each station. The set of $q$ operators that are applied in a given moment is named a *configuration*. F.ex. for $q=3$, $\sigma_l^{(1)} \otimes \sigma_l^{(2)} \otimes \sigma_r^{(3)} \equiv llr$ is a possible *configuration*. It means that $\sigma_l$ is applied in stations 1 and 2, and $\sigma_r$ in station 3.

The observation of a qubit gives the result +1 or -1 in each station. The *total result* of an observation performed on $|\phi_{(q)}\rangle$ is the product of the results obtained at each station. The eigenvalues (of *total results*) of $|\phi_{(q)}\rangle$ are $(i)^{R-1}$ for R odd, where R is the number of letters $r$ in the *configuration*. For *configurations* with R even, $|\phi_{(q)}\rangle$ is not an eigenvector. The *configurations* for which $|\phi_{(q)}\rangle$ is (is not) eigenvector are named *words* (*strings*). For *strings*, the *total result* is (+1) or (-1) with equal probability.

F.ex. for $q = 3$, the *words* are *llr*, *lrl*, *rll* (eigenvalue +1) and *rrr* (eigenvalue -1). The remaining 4 *configurations*: *lrr*, *rlr*, *rrl*, *lll* are *strings*. A simple hidden variables model uses 2×3 matrices to determine the results of the observations at each station. It is able to reproduce the QM predictions for 3 of the 4 *words* and for all the *strings* [6]. The remaining *word* (the one whose QM prediction the carried matrix is unable to reproduce) is called a *bad word*. The matrix carried by the trio and the *configuration* the trio finds at the stations are, by hypothesis, uncorrelated. Therefore, the probability that the model is unable to reproduce the QM predictions (in other terms: the probability that the matrix carried by the trio finds its *bad word* at the stations) is 1/8, not 1. Hence, Local Realism cannot be single shot disproved, as it is often believed. It can only be statistically disproved. In an ideal setup it is necessary to observe at least 35 trios to disprove Local Realism with a certainty >99%. Instead, *what can be single shot disproved is QM*. F.ex., it suffices to observe the *total result* -1 for the *word llr* (for which the eigenvalue is +1) just once, and QM is disproved. This is valid for all values of $q$.

Of course, single-shot disproval of QM can occur in an ideal setup only. In a minimal approximation to reality, some space for "errors" must be allowed. Regardless the cause of the errors, what happens at each station is uncorrelated from what happens at the other stations, hence, a probability $\varepsilon$ (<<1) of error per station must be defined. The probability of observing a *total result* that deviates from QM predictions (what is called here a *failure*) is then:

$$P_{QM\,failure} = \frac{1}{2}\sum_{j\,odd}^{q}\binom{q}{j}\varepsilon^j(1-\varepsilon)^{q-j} =$$
$$= ¼ - ¼ (1-2\varepsilon)^q \qquad (2)$$

The factor ½ in the first line is because *failures* do not occur for *strings* (that appear half the times, for the operator at each station is chosen in an uncorrelated way) and the sum is over odd values of $j$ because an even number of errors leaves the sign of the *total result* unchanged (thus they produce no *failures*).

On the other hand, a model holding to Local Realism must produce a number of *failures* which is given (in the absence of loopholes or conspiratorial behavior) by the rate of *bad words* in the total number of *configurations*. The number $B(q)$ of *bad words* increases with $q$. From Mermin's inequality [5] it is derived [6] that:

$$B(q) \geq 2^{q-2} - 2^{(q-2)/2} \text{ for } q \text{ even,} \qquad (3)$$
$$2^{q-2} - 2^{(q-3)/2} \text{ for } q \text{ odd,}$$

where the equality holds for deterministic classical models (that is: models where the results of the observations at each station are determined by the carried hidden variable, as in the matrices' model mentioned before). The probability of *failures* is then:

$$P_{Classical\,failure} = B(q) / 2^q \to ¼ \quad \text{if } q>>1 \qquad (4)$$

from eqs.(2) and (4):

$$P_{Classical\,failure} - P_{QM\,failure} = ¼ (1-2\varepsilon)^q \to 0 \quad \text{if } q>>1 \qquad (5)$$

Hence, the number of *failures* caused by the tolerance allowed to avoid a single shot disproval of QM, and the number caused by the limitations of the best classical model, are indistinguishable in the limit $q>>1$. Assuming *(i)* and *(ii)* valid, eq.(5) means that the classical limit is reached by only increasing the number of elementary particles in the system.

To get an idea of the situation for a macroscopic system, consider a (Schrödinger's) cat. The cat weights 4 Kg and is made mostly of water, then: $q \approx 4\times10^{27}$. Suppose all particles are GHZ-entangled (in principle at least, the cat's state farthest from classical). In order to observe a difference $>10^{-2}$ in the rate of *failures* between classical and QM predictions it is necessary to get $\varepsilon < 6\times10^{-28}$, an unreachable small value. Recall that supposing $\varepsilon$ identically equal to zero is both unrealistic and undesirable, for it would allow a single shot disproval of QM.

The result that the classical limit is reached by merely increasing the number of particles may be surprising but, actually, there were antecedents indicating that it might be so. It was shown [6] that disproving Local Realism with GHZ states was increasingly difficult if $q$ was increased (up to $q = 8$), even if the difficulty in preparing the state was not taken into account. This was the consequence of a hidden variable model that exploited the predictability loophole. In this case, the cause of the surprising result was that the number of *words* increased faster (with $q$) than the number of *bad words* that could be reached from the *configuration* that existed when the particles were emitted. Another antecedent is that entanglement swapping (which involves 4 particles) can be classically described, by exploiting the detection loophole, for conditions less restrictive than for the usual Bell's experiment with 2 particles [7]. It is worth remarking that eq.5 is derived here without loopholes or conspiratorial behavior of any kind. Even the detectors' efficiencies (a caveat stated in [5]) are assumed perfect. The surprising result here is the consequence that *both* $P_{QM}$ and $P_{Classical} \to ¼$ in the limit $q >>1$. This result is consistent with the view that all infinite-dimensional systems are classical [8].

An objection: recall that GHZ states of qubits are not the only possible form of entanglement among many particles (see f.ex. [9,10]). Assumption *(ii)* may be wrong. But, if it is demonstrated right, the result for GHZ will include the others. Another objection is that the set of operators used in Mermin's inequality are conceivably not the only ones able to refute Local Realism with GHZ states. In fact, this set was introduced in [5] as an example of just *one* way to do it. This set has become habitual, but it was never claimed to be the only possible one, or the best. The results may be different for another set of operators. These two objections entail lines of research important by themselves, and of a complexity beyond the scope of this short paper. Hopefully, the result presented here will encourage the activity along these lines. Assuming the objections are favorably elucidated, stating the classical realm is reached when the number of elementary particles composing the system is macroscopic (no matter their entanglement), appears as the simplest way to express the principle of correspondence.

**Acknowledgments.**

Many thanks to Prof. Federico Holik for a critical reading of the first version of this manuscript. This work received support from the grants N62909-18-1-2021 Office of Naval Research Global (USA), and PIP 2017-027C CONICET (Argentina).

**References.**

[1] See f.ex. the contributions in "Non Hermitian Hamiltonians in Quantum Physics"; *Selected Contributions*

*from the 15th International Conference on Non-Hermitian Hamiltonians in Quantum Physics*, Palermo, Italy, 18–23 May 2015, F.Bagarello, R.Passante and C.Trapani Ed., *Springer Proceedings in Physics* **184** (2016).

[2] Ghirardi, A.Rimini and T.Weber, "Unified dynamics for microscopic and macroscopic systems", *Phys. Rev. D.* **34** p.470 (1986).

[3] H.Everett, "Relative State Formulation of Quantum Mechanics", *Rev.Mod.Phys.* **29** p.454 (1957).

[4] A.Bassi *et al.*, "Models of wave-function collapse, underlying theories and experimental tests"; *Rev.Mod.Phys.* **85** p.471 (2013).

[5] N. David Mermin, "Extreme quantum entanglement in a superposition of macroscopically distinct states", *Phys. Rev. Lett.* **65** p.1838 (1990).

[6] A.Hnilo, "On testing objective local theories by using GHZ states" *Found.Phys.* **24** p.139 (1994).

[7] N.Gisin and B.Gisin, "A local variable model for entanglement swapping exploiting the detection loophole", *Phys. Lett.* A **297** p.279 (2002).

[8] K.Hepp, "Quantum theory of measurement and macroscopic observables", *Helv.Phys.Acta* **45** p.237 (1972).

[9] O.Gühne *et al.*, "Bell inequalities for graph states", *Phys. Rev. Lett.* **95**, 120405 (2005).

[10] V.Scarani *et al.*, "Nonlocality of cluster states of qubits", *Phys. Rev. A* **71**, 042325 (2005).